\documentclass[useAMS,usenatbib]{mnras}

\usepackage{graphicx}
\usepackage{epstopdf}
\usepackage{amsmath,amssymb,esint}

\title[Retrograde Accretion Disks in Be-HMXBs]{Retrograde Accretion Disks in High-Mass Be/X-ray Binaries}

\author[D. M. Christodoulou et al.]{D. M. Christodoulou,$^{1,2}$\thanks{E-mail: dimitris\_christodoulou@uml.edu} 
S. G. T. Laycock,$^{1,3}$\thanks{E-mail: silas\_laycock@uml.edu}
and D. Kazanas$^{4}$\thanks{E-mail: demos.kazanas@nasa.gov}
\\
$^{1}$Lowell Center for Space Science and Technology, University of Massachusetts Lowell, Lowell, MA, 01854, USA\\
$^{2}$Department of Mathematical Sciences, University of Massachusetts Lowell, Lowell, MA, 01854, USA\\
$^{3}$Department of Physics \& Applied Physics, University of Massachusetts Lowell, Lowell, MA, 01854, USA\\
$^{4}$NASA Goddard Space Flight Center, Laboratory for High-Energy Astrophysics, Code 663, Greenbelt, MD 20771, USA\\
}

\begin{document}

\def\gsim{\mathrel{\raise.5ex\hbox{$>$}\mkern-14mu
                \lower0.6ex\hbox{$\sim$}}}

\def\lsim{\mathrel{\raise.3ex\hbox{$<$}\mkern-14mu
               \lower0.6ex\hbox{$\sim$}}}

\pagerange{\pageref{firstpage}--\pageref{lastpage}} \pubyear{2016}

\maketitle

\label{firstpage}

\begin{abstract}

We have compiled a comprehensive library of all X-ray observations of Magellanic pulsars carried out by {\it XMM-Newton}, {\it Chandra}, and {\it RXTE} in the period 1997-2014. In this work, we use the data from 53 high-mass Be/X-ray binaries in the Small Magellanic Cloud to demonstrate that the distribution of spin-period derivatives vs. spin periods of spinning-down pulsars is not at all different than that of the accreting spinning-up pulsars. The inescapable conclusion is that the up and down samples were drawn from the same continuous parent population, therefore Be/X-ray pulsars that are spinning down over periods spanning 18 years are in fact accreting from retrograde disks. The presence of prograde and retrograde disks in roughly equal numbers supports a new evolutionary scenario for Be/X-ray pulsars in their spin period-period derivative diagram.

\end{abstract}




\section{Introduction}\label{intro}

We have produced a comprehensive library of all X-ray observations of Magellanic pulsars carried out by the {\it XMM-Newton}, {\it Chandra}, and {\it RXTE} telescopes in the period 1997-2014. The Small Magellanic Cloud (SMC) contains 63 pulsars in 
high-mass Be/X-ray binaries (HMXBs) \citep{coe15,hab16} for which all X-ray data have been released for public use \citep{yang17}. In this Letter, we focus specifically on 53 of these pulsars for which we can argue that many are not in spin equilibrium over the entire observational period spanning 18 years. For these pulsars, \cite{yang17} carried out a linear regression of their spin periods $P_S$ on timescales much longer than the orbital periods and they measured their long-term period derivatives $\dot{P_S}$. (For the remaining pulsars in the sample, no pulsations were measured.) Owing to the long observational timescales, these measurements are as robust as currently possible, and they remain unaffected by minor reversals of the $\dot{P_S}$ values in the short term and by periodic variations caused by orbital doppler motions. The results show that nearly half of these objects are consistently spinning down in the long term which is surprising because we do not know of any physical mechanism as efficient as accretion that would instead slow down these pulsars. This argument was made by \cite{chr17} in the context of ultraluminous X-ray pulsars (all three of which are spinning up) and it obviously extends to Be/X-ray pulsars as well.

In order to study spin changes more closely, we separate the \cite{yang17} data into two samples, one with pulsars spinning up and another with pulsars spinning down (the ``up'' and ``down'' samples); and we compare statistically the two-dimensional (2-D) distributions of their dynamical properties (spin-period derivative vs. spin period) in the two samples (\S~\ref{data}). We find that the two samples are statistically quite similar. Finally, we present and discuss our conclusions in \S~\ref{conclusions}, where we argue that the results support the presence of retrograde accretion disks in spinning-down pulsars and the presence of moderate magnetic fields in the compact objects of the entire class. Based on these conclusions, we also describe an evolutionary path for the pulsars in the $P_S-\dot{P_S}$ diagram.

\section{Data Analysis}\label{data}

\begin{figure}
\begin{center}
    \leavevmode
      \includegraphics[trim=0 0.1cm 0 0.1cm, clip, angle=0,width=9 cm]{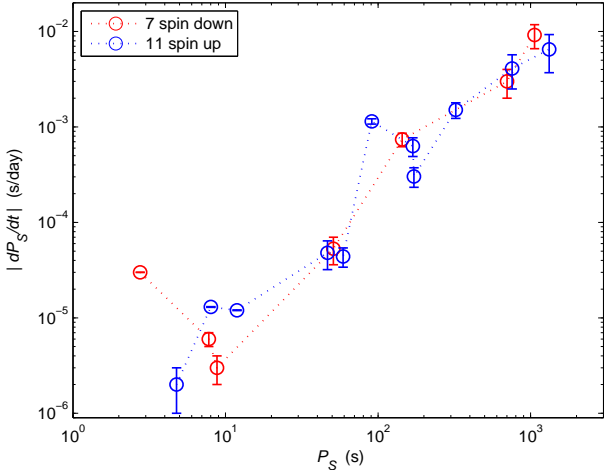}
\caption{The absolute value of period derivative ($\dot{P_S}=dP_S/dt$) is plotted against spin period $P_S$ for the SMC pulsars that are spinning up or down at a significance level of 2$\sigma$ or better \citep{yang17}. There is no difference between the distributions of $|\dot{P_S}|$ vs. $P_S$ in the two samples and this indicates that the long-term changes in spin are caused by accretion in both types of pulsars.
\label{fig1}}
  \end{center}
\end{figure}

The data that we use in this analysis were published by \cite{yang17}, where the pulsars were divided into groups depending on the sign and magnitude of the period derivatives $\dot{P_S}=dP_S/dt$ and their errors. Using the measured errors in $\dot{P_S}$, a conservative $2\sigma$ criterion was initially adopted to separate the objects into those that appear to be close to spin equilibrium and those that are clearly spinning up or down. \cite{yang17} then found that $N_{up}=11$ pulsars are spinning up and $N_{down}=7$ pulsars are spinning down. These two samples are plotted in Fig.~\ref{fig1} using blue and red open circles connected by dotted line segments. The $\dot{P_S}$ error bars are also plotted; owing to the stringent $2\sigma$ criterion, all of them are small.

\begin{figure}
\begin{center}
    \leavevmode
      \includegraphics[trim=0 0.1cm 0 0.1cm, clip, angle=0,width=9 cm]{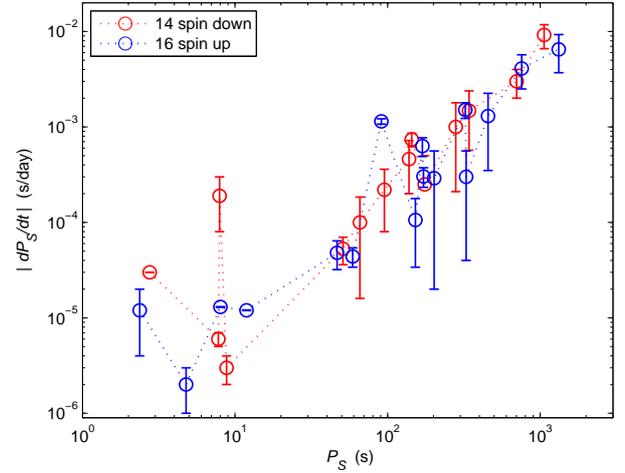}
\caption{As in Fig.~\ref{fig1}, but for the SMC pulsars that are spinning up or down at the less conservative significance level of 1$\sigma$ or better.
\label{fig2}}
  \end{center}
\end{figure}

%
\begin{table*}
\caption{Statistical Tests for Two-Dimensional (2-D) Data Samples}
\label{t1}
\begin{tabular}{lclc}
\hline
\multicolumn{4}{c}{$2\sigma$ Samples ($N_{up}=11$, $N_{down}=7$)}\\
\hline
Test & References & Statistic & $p$-value \\
\hline
Kolmogorov-Smirnov 2-D  & 1-3 &    $D$=0.2424                &   0.9533 \\
Minimum Energy (SR)       & 4 &   $E_{nm}$=84.1817      &   0.9800 \\ 
Minimum Energy (AZ)       & 5 &     $E_{nm}=  -0.2472$  &   0.9300 \\
Hotelling's $T^2$             &  6 &   $T^2$=1.1707             &   0.5888 \\

\hline
\multicolumn{4}{c}{$1\sigma$ Samples ($N_{up}=16$, $N_{down}=14$)}\\
\hline
Test & References & Statistic & $p$-value \\
\hline
Kolmogorov-Smirnov 2-D  & 1-3 &   $D$=0.2404               &    0.7736 \\
Minimum Energy (SR)       & 4 &   $E_{nm}$=50.2773    &    0.9860 \\
Minimum Energy (AZ)       & 5 &   $E_{nm}= -0.1688$   &    0.9920 \\
Hotelling's $T^2$             &  6 &   $T^2$=1.8852           &    0.4149 \\

\hline
\end{tabular}
\\
Ref. Key---(1)~\cite{pea83}; (2)~\cite{fas87}; (3)~\cite{pre92}; (4)~\cite{sze14}; (5)~\cite{asl05}; (6)~\cite{mar79}.

\end{table*}

\cite{yang17} finally settled to using a $1.5\sigma$ criterion for the data.
In our analysis, we filter the same data according to a less stringent $1\sigma$ level of significance. In this case, we find that $N_{up}=16$ pulsars are spinning up and $N_{down}=14$ pulsars are spinning down. These larger samples are plotted in Fig.~\ref{fig2} using blue and red open circles connected by dotted line segments. The $\dot{P_S}$ error bars are also plotted; the error bars of the additional points relative to those in Fig.~\ref{fig1} are large and extend well outside the open circles used in the figure. Despite the larger errors in Fig.~\ref{fig2}, we observe in both figures that the distributions of $|\dot{P_S}|(P_S)$ are quite similar for the up and down samples (blue and red circles, respectively). 
In fact, there appears to be a trend in all four samples at confidence levels of 90-95\%, viz. 
\begin{equation}
\log|\dot{P_S}|\approx k\cdot \log{P_S} - 6\, , ~~ k=1.1-1.4 \, , 
\end{equation}
where $\dot{P_S}$ is expressed in seconds/day and $P_S$ in seconds. 

A 2-D statistical analysis confirms the results from visual inspection without a doubt: We adopted the null hypothesis that, in each case, the up and down samples were drawn from the same continuous population $|\dot{P_S}|(P_S)$, and we applied four independent statistical tests\footnote{We used a suite of multidimensional two-sample tests that was written by \cite{lau14} and that is distributed freely under the GNU General Public License of the Free Software Foundation.} in which the null hypothesis cannot be rejected at any reasonable level of significance. The results are summarized in Table~\ref{t1}. In all cases, the $p$-values are extremely large indicating that the samples were drawn from the same continuous population. The smallest $p$-value ($p=0.4149$) occurs for Hotelling's $T^2$ test \citep{mar79} in the $1\sigma$ samples that are supposed to be less rigorous than the $2\sigma$ samples; but even in this case, the $p$-value is so large that one cannot reject the null hypothesis at any reasonable level of significance.

The results do not change at all in the less constrained one-dimensional (1-D) comparison of the two $|\dot{P_S}|$ samples irrespective of spin periods: The rigorous 1-D Kolmogorov-Smirnov test \citep{mas51,mar03} confirms the null hypothesis that the up and down $|\dot{P_S}|$ samples were drawn from the same continuous population for any reasonable level of significance. The $p$-values are very large, 0.9911 for the $2\sigma$ data and 0.9821 for the $1\sigma$ data.

The remarkable strength of the null hypothesis in the above statistical tests prompted an examination of the raw data in \cite{yang17}, where a total 53 pulsars are listed as having nonzero period derivatives irrespective of significance. This data set is expected to be contaminated by 23 additional pulsars that are very close to spin equilibrium and their measured $|\dot{P_S}|$ values are too small to be meaningful. We ran the same statistical tests on the raw data ($N_{up}=27$ and $N_{down}=26$) and, once again, we could not reject the null hypothesis at any reasonable level of significance. This is because, in this case, the elevated noise is restricted to small $|\dot{P_S}|$ values and occurs in both samples. Thus, our result appears to be statistically robust regardless of the chosen $\sigma$-criterion or the level of significance for the null hypothesis.

\section{Conclusions}\label{conclusions}

\subsection{Retrograde Accretion Disks}

We have studied the statistical properties of two samples of Be-HMXB pulsars in the Small Magellanic Cloud, one with the pulsars spinning up due to accretion over the past 18 years, and another with the pulsars spinning down over the same period.
We found absolutely no difference between the two samples: four independent statistical tests \citep[Table~\ref{t1} and][]{lau14} show that both samples were obtained from the same continuous parent population. We believe that this constitutes evidence that the physical processes are the same in the two groups of pulsars, therefore spinning-down pulsars are also undergoing accretion and their spin periods increase only because the accreted angular momentum is obtained from retrograde disks whose rotation vectors are antiparallel to the spin vectors of the neutron stars.

Retrograde accretion disks have been previously proposed for a few stellar-mass black holes \citep{zha97,rao12,rei13,mid14,mor14} and for some supemassive black holes in Active Galactic Nuclei \citep{gar10,cow12,gar13}, but not for neutron stars in Be/X-ray binaries. A retrograde fossil accretion disk has also been proposed for the anomalous single pulsar CXOU J010043.1-721134 in the SMC \citep[\S3.4 in][and references therein]{chr16}. The results of our study indicate that, at this time, there appears to be a larger number ($\approx 50\%$) of retrograde disks around Be/X-ray pulsars than around black holes of all sizes. This result is not really in disagreement with the conclusions from extensive population-synthesis studies that find a smaller percentage ($\sim 5\%$) of retrograde accretion disks in X-ray binaries \citep[e.g.,][]{fra10} because these studies were conducted for Roche-lobe filling supergiant companions that give rise to repeated, short-lived, wind-fed accretion events around compact objects in which the direction of rotation of the accreted gas may be changing from one event to the next \citep[][and references therein]{bil97,van98}.

\subsection{Magnetic Fields}

Another conclusion that can be drawn from the above results concerns the magnitudes of the surface magnetic fields in the Be-HMXB pulsars: it is improbable that the slowdown of the pulsars in the down samples is caused by strong, magnetar-type \citep{woo06,ola14} magnetic fields. If that were the case, then such strong magnetic fields would be able to also counter and slow down the spinup of the accreting pulsars in the up samples, thereby changing the distributions of $|\dot{P_S}|(P_S)$ of the up and down samples. Since no such statistical differences are observed between the two samples, we conclude that there cannot be two different mechanisms changing the spin periods and that the magnetic fields are moderate ($B\sim 0.3-10$~TG) in the entire class of Be-HMXB pulsars, as was also found in our two previous studies \citep{chr16,chr17}.

\begin{figure}
\begin{center}
    \leavevmode
      \includegraphics[trim=0 0.1cm 0 0.1cm, clip, angle=0,width=9 cm]{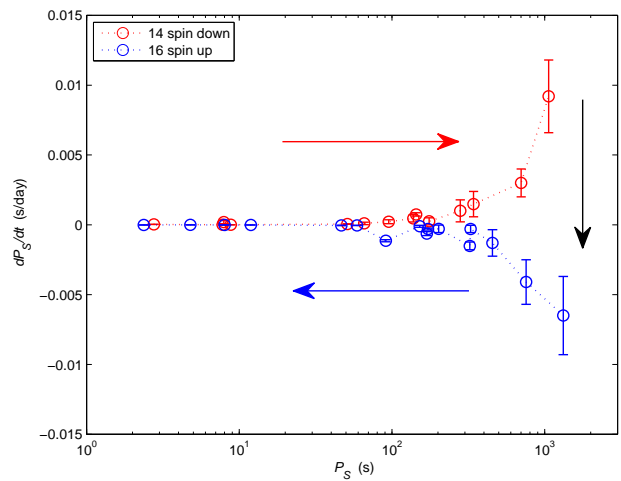}
\caption{The period derivative ($\dot{P_S}=dP_S/dt$) is plotted on a linear scale against spin period $P_S$ for the $1\sigma$ samples of Fig.~\ref{fig2}. The arrows show the direction in which the spin periods evolve over time.
\label{fig3}}
  \end{center}
\end{figure}

\subsection{Spin Evolution}

The marked similarity between the up and down samples is an invitation to combine the two samples into a larger data set. A linear regression to the combined $2\sigma$ data set gives
\begin{equation}
\log|\dot{P_S}|= 1.31(\pm 0.12)\cdot \log{P_S} - 6.21(\pm 0.23) \, , 
\end{equation}
where $\dot{P_S}$ is expressed in seconds/day and $P_S$ in seconds. The correlation coefficient is $r^2=0.884$ and the $p=0.058$ value is close to the 95\% confidence level. The results are similar for the combined $1\sigma$ data set:
\begin{equation}
\log|\dot{P_S}|= 1.12(\pm 0.11)\cdot \log{P_S} - 5.85(\pm 0.21) \, , 
\end{equation}
where $r^2=0.802$ and $p=0.060$. As expected, the correlation is now degraded due to the elevated noise in the $1\sigma$ data, but the statistic holds firm near the 95\% confidence level. The implication from these results is that spinning up and down pulsars cannot be separated by their $P_S$ values; that is, spinning up/down pulsars do not have predominantly shorter/longer spin periods, as one might have expected; and this observation lends strong support for a surprising evolutionary scenario that cannot be detected in Figs.~\ref{fig1} or~\ref{fig2} where absolute values are plotted. In order to describe this scenario, we plot in Fig.~\ref{fig3} the $1\sigma$ samples with $\dot{P_S}$ shown on a linear scale. The $2\sigma$ data show the exact same features, albeit with fewer points.

We observe in Fig.~\ref{fig3} that only long-period pulsars show $\dot{P_S}$ values that are substantially different than zero and there is a congested area around $P_S = 140$-175~s. Long-period pulsars spinning up (blue circles) necessarily evolve in the direction of the blue arrow but their rates are slowed down considerably after they cross the congested area. Below $P_S\approx 100$~s, these pulsars are effectively close to spin equilibrium, although they may continue their spin-up very slowly. In contrast, short-period pulsars spinning down (red circles) necessarily evolve in the direction of the red arrow but they do so very slowly until they cross the congested area, to the right of which they spin down much faster.  In both cases, accretion does not appear to be too efficient in changing the spin periods of the short-period pulsars, thus many accretion events are required for these pulsars to make progress in either direction. This entire evolutionary scenario can be understood if the accretion disks can reverse their direction of rotation when the pulsars are near spin equilibrium (e.g., a new accretion phase begins but the inflow has its rotation opposite to that of the preceding phases). Then these pulsars may evolve back and forth near the $\dot{P_S}=0$ line --- but, eventually, sustained accretion in one direction may send some of them in the direction of the red arrow. In fact, hints of such reversals have been seen in the \cite{yang17} data which would suggest that reversals can occur over timescales of only $\gtrsim$20~years.

Another reversal of the rotation of the inflow at very long spin periods closes the loop in Fig.~\ref{fig3} and predicts the fate of long-period pulsars with $P_S\gtrsim 1000$~s that continue to spin down: these objects cannot return from the path that brought them to long periods (red arrow). With their angular-momentum contents reduced so much ($\sim$1000 times), the next major ``retrograde'' accretion event (or events) should have no difficulty reversing their spins, facilitating the transition depicted by the black arrow in Fig.~\ref{fig3}. This transition appears to proceed fast because no pulsars are seen at the longest periods with their $\dot{P_S}$ values near zero.

It is interesting to note that in the current four samples ($1\sigma$ and $2\sigma$, up and down), the pulsars are separated above and below the characteristic spin period of 100~s in equal numbers. This observation also supports the evolutionary scenario discussed above: both groups have to slow down (speed up) their spin evolution at short (long) periods, and both groups have to wait for the same agent (a sequence of consecutive ``retrograde'' accretion events) that will initiate the process that eventually leads to a spin reversal.

\section*{Acknowledgments}
We are obliged to an anonymous referee for critical comments and suggestions that led to a number of improvements in the paper. DMC and SGTL were supported by NASA grant NNX14-AF77G. DK was supported by a NASA ADAP grant.

\label{lastpage}

\end{document}